%% file: main.tex
 \documentclass[final,5p,times,twocolumn]{elsarticle}

\usepackage{amssymb}
 \usepackage{amsthm}
\newtheorem{definition}{Definition}
\newcommand{\qop}[1]{\textsf{#1}}
\newcommand{\perf}{\mathit{perf}}

\journal{Neurocomputing}

\usepackage[ruled,linesnumbered,lined]{algorithm2e}
\usepackage{amsmath}
\usepackage[utf8]{inputenc}
\usepackage{tikz}

\input{Qcircuit}

\begin{document}

\begin{frontmatter}



\title{Weightless neural network parameters and architecture selection in a quantum computer}


\author[ufrpe,ufpe]{Adenilton J. da Silva$^\dag$}
\author[ufrpe]{Wilson R. de Oliveira}
\author[ufpe]{Teresa B. Ludermir}
\address[ufrpe]{Departamento de Estat\'{i}stica e Inform\'{a}tica \\ Universidade Federal Rural de Pernambuco, Brazil}
\address[ufpe]{Centro de Inform\'{a}tica \\ Universidade Federal de Pernambuco, Brazil}

\begin{abstract}
Training artificial neural networks requires a tedious empirical evaluation to determine a suitable neural network architecture. To avoid this empirical process several techniques have been proposed to automatise the architecture selection process.  In this paper, we propose a method to perform parameter and architecture selection for a quantum weightless neural network (qWNN). The architecture selection is performed through the learning procedure of a qWNN with a learning algorithm that uses the principle of quantum superposition and a non-linear quantum operator. The main advantage of the proposed method  is that it performs a global search in the space of qWNN architecture and parameters rather than a local search.

\end{abstract}

\begin{keyword}


 Quantum neural networks \sep quantum weightless neural networks \sep quantum learning 
\sep architecture selection

\end{keyword}

\end{frontmatter}



\section{Introduction}

 The exponential reduction of computers' components known as Moore’s law took 
computation from the classical physical domain to the quantum physics. The idea of quantum computation was initially proposed in \cite{Feynman}, where Feynman states that quantum computers can simulate quantum physical systems 
exponentially faster than classical computers. Some quantum algorithms also 
overcome the best knew classical algorithms; the most famous examples being the Shor’s factoring algorithm \cite{shor:97} that is exponentially faster than the best know classical algorithm and the Grover’s search algorithm \cite{grover:96} with quadratic gain in relation to the best classical algorithm for unordered search. It is true that quantum computers are not yet a reality, but there has been an explosion of investment in quantum computing, the result of which are numerous proposals for quantum computers and the general belief which soon they will be realised.  The use of an adiabatic quantum system with 84 quantum bits is reported in~\cite{Bian2013}  and in \cite{Monz2011} is reported the creation of a quantum system with 14 quantum bits. Empirical evaluations of ideas presented in this work for real problems require a quantum computer with capacity to manipulate hundreds of qubits which is impossible with current technology.

 One of the main characteristics of quantum computation is the quantum parallelism that for some problems allows quantum algorithms to have a speedup in relation to the classical algorithms. With quantum parallelism is possible to calculate all possible $2^n$ values of a $n-$ary Boolean function in a single query. However, we cannot visualise these outputs 
directly. A quantum measurement is necessary and it returns probabilistically only a more restrict value. The quantum algorithm design problem is then to perform quantum operations to increase the probability of the desired output. 

 Designing quantum algorithms is not an intuitive task. Attempts to 
bring quantum computing power to a greater range of problems are the development of quantum machine-learning algorithms as decision trees \cite{Farhi1998}, evolutionary algorithms \cite{Malossini2008} and artificial 
neural networks \cite{panella:09,Altaisky,oliveira:08,ventura:04,Behrman,daSilva:12,Narayanan,Oliveira2009,Liu2013}. In this paper, we are concerned in the field of quantum weightless neural networks. 

Weightless neural networks (WNN) are not the most used model of artificial neural networks. WNN have been proposed by Aleksander~\cite{Aleksander1966} as engineering tools to perform pattern classification. Applications of WNN are described in several works ~\cite{Staffa2014a, Carvalho2014, Cardoso2014,esann:2014:tutorial} and quantum versions of WNN have been proposed in~\cite{oliveira:08,Oliveira2009,daSilva:12}.

The idea of quantum neural computation has been proposed in the nineties~\cite{kak:95}, since then
several models of quantum neural networks have been proposed. For instance, quantum weightless neural networks 
\cite{daSilva:12}, neural networks with quantum architecture \cite{panella:09} 
and a simple quantum neural network \cite{ventura:04}. In all these works 
\cite{daSilva:12,panella:09,ventura:04} a quantum neural network configuration 
is represented by a string of qubits and quantum learning algorithms are 
proposed within a common framework. The main idea of the learning algorithms in 
\cite{daSilva:12,panella:09,ventura:04} is to present input data to all possible 
neural networks for a given architecture in superposition and perform a quantum 
search in the resulting superposition. The objective of this paper is to 
generalise this idea to allow architecture selection through the training of a quantum 
weightless neural network. To achieve this objective we use a quantum weightless 
neural network that stores representations of weightless neural networks with different architectures in its memory 
positions and we define a quantum learning algorithm using the non-linear operator proposed 
in \cite{PhysRevLett.81.3992} and the measurement and feedback strategy~\cite{Gammelmark:09}.

 Selection of a neural network architecture is an important task in  neural networks applications.
 Normally this task requires a lot of empirical evaluation performed by an expert. 
 To avoid the tedious empirical evaluation process and help inexperienced users some algorithms have been proposed to perform automatic selection of neural networks architecture. Techniques such as  meta-learning~\cite{Abraham20041} and 
evolutionary computation~\cite{Almeida2010} have been used for architecture selection. 

In this paper, we show how to use a quantum weightless neural network with a non-linear quantum-learning algorithm to find a quantum neural network architecture and parameters with a desired performance. The proposed algorithm uses quantum superposition principle and a non-linear quantum operator. The proposed algorithm performs a global search in architecture and parameters space and its computational time is polynomial in relation to the number of training patterns, architectures and quantum weightless network memory size.

     The rest of the paper is organised as follows. Section 2 presents basics concepts on quantum computation such as quantum bits, operators, measure and parallelism. Section 
3 presents the concept of weightless neural networks, quantum neural networks and quantum weightless neural networks. 
Section 4 describes a quantum learning algorithm for weightless neural networks 
and how to apply this learning algorithm to perform architecture selection.
Finally, Section 5 is the conclusion.

\section{Quantum computing}
Deep knowledge of classical physics is not required for designing classical algorithms. In the same vein, the development of quantum algorithms does not require a deep knowledge of quantum physics and there are several books~\cite{nielsen:00, Hirvensalo2003,mermin2007quantum} that follow this approach by introducing only the strictly necessary knowledge of quantum physics for the understanding of quantum computing. In order to create a  self-contained text a brief introduction to quantum computing is presented. 

The state of a quantum computer with $n$ quantum bits (or \emph{qubits}) can be mathematically represented by a unit vector of an $2^n$-dimensional complex vector space with inner product. For instance, one single qubit can be represented in the vector space $\mathbb{C}^2$ as described in Equation~\eqref{eq:qubit},
\begin{equation}
\ket{\psi}=\alpha\ket{0}+\beta\ket{1}
\label{eq:qubit}
\end{equation}
where $\alpha, \beta \in \mathbb{C}$, $\left|\alpha\right|^2+\left|\beta\right|^2 = 1$ and $\ket{0}$ and 
$\ket{1}$ are the vectors described in Equation~\eqref{eq:canbasis}\footnote{We
could have used any other orthonormal basis but in quantum computing the canonical basis also called \emph{computational basis} is the most employed.}.
\begin{equation}
\ket{0} = \begin{bmatrix} 1 \\ 0 \end{bmatrix} \mbox{ and } \ket{1} = \begin{bmatrix} 0 \\ 1 \end{bmatrix}
\label{eq:canbasis}
\end{equation}
One qubit in a $n$-dimensional quantum system is represented by the $2^n$-dimensional vector space as described in Equation~\eqref{eq:qubits},
\begin{equation}
\sum_{i=0}^{2^n-1}\alpha_i \ket{\psi_i}
\label{eq:qubits}
\end{equation}
where the sum of the squared modulus of the amplitude $\sum_i |\alpha_i|^2$ is equal to one 
and the set $\{\ket{\psi_0}, \ket{\psi_1}, \cdots, \ket{\psi_{2^n-1}} \}$ is an orthonormal basis of $\mathbb{C}^{2^n}$.

A \emph{Quantum operator} in a quantum system with $n$ qubits is an unitary
operator\footnote{An operator (or matrix, for the finite dimensional case once
fixed a basis) $A$ is \emph{unitary} if $AA^\dagger=A^\dagger A=I$ where
$A^\dagger$ is the complex conjugate of the transpose of $A$} in the vector
space $\mathbb{C}^{2^n}$. Let $U$ be an unitary operator over $\mathbb{C}^{2^n}$
and $\ket{\psi_{t_1}}$ the state of the quantum system. After applying the quantum
operator $U$ the system will be in the state $\ket{\psi_{t_2}}$ described in
Equation~\eqref{eq:evolution}.
\begin{equation}
\ket{\psi_{t_2}} = U\ket{\psi_{t_1}}
\label{eq:evolution}
\end{equation}
In the computational basis, the matrix representation of the quantum operators the \emph{not operator} $X$ and the \emph{Hadamard operator} $H$ over one qubit are described in Equation~\eqref{eq:exop}. 
\begin{equation}
X = \begin{bmatrix}
0 & 1 \\
1 & 0
\end{bmatrix} \mbox{ and } 
H =\frac{1}{\sqrt{2}} \begin{bmatrix}
1 & 1 \\
1 & -1
\end{bmatrix}
\label{eq:exop}
\end{equation}
$X$ acts on the computational basis vectors as a not operator ($X\ket{0}=\ket{1}$ and $X\ket{1}=\ket{0}$)  and $H$ applied to a state in the computational basis creates a ``uniform"  superposition (or linear combination) of the two basis:
\begin{equation}
\begin{split}
H\ket{0} = \frac{1}{\sqrt{2}}(\ket{0} + \ket{1})\\
H\ket{1} = \frac{1}{\sqrt{2}}(\ket{0} - \ket{1})
\end{split}
\label{eq:exH1}
\end{equation}
both represent a state which is $\ket{0}$ with probability $\frac{1}{2}$ and $\ket{1}$ with probability $\frac{1}{2}$, and can be thought of a state which is both $\ket{0}$ and $\ket{1}$. That is why one says that a qubit is able to ``store" two classical bits simultaneously. This scale up exponentially with the number of qubits $n$, $H\ket{0}\otimes \cdots \otimes H\ket{0} = H^{\otimes n}\ket{0\cdots 0}$, with $0\cdots 0$ being a sequence of $n$ $0$'s, is the superposition of all $2^n$ possibles $n$-qubits.  Equation~\eqref{eq:exH} shows the result for $n=2$, where $H^{\otimes 2} = H\otimes H$:
\begin{equation}
\begin{split}
H^{\otimes 2} \ket{0}\ket{0} = \frac{1}{2}\left(\ket{0} + \ket{1}\right)\otimes \left(\ket{0} + \ket{1}\right)
= \\ \frac{1}{2}(\ket{0}\ket{0} + \ket{0}\ket{1} + \ket{1}\ket{0} + \ket{1}\ket{1})
\end{split}
\label{eq:exH}
\end{equation}

\emph{Quantum parallelism} is one of the main properties of quantum computation and it is used in the majority of quantum algorithms. Let $U_f$ be a quantum operator with action described in Equation~\eqref{eq:uf},
\begin{equation}
U_f\ket{x,c} = \ket{x, c \oplus f(x)}
\label{eq:uf}
\end{equation}
where $f:B^m \rightarrow B^n$ is a Boolean function. Applying this operator in a state in superposition $\sum_i\ket{x_i,0}$ the value of $x_i$ will be calculated for all $i$ in a single quantum operation, as described in Equation~\eqref{eq:parallelism}.
\begin{equation}
U_f\left(\sum_{i}\ket{x_i,0}\right) = \sum_{i}U_f \ket{x_i,0} = \sum_{i}\ket{x_i,f(x_i)}
\label{eq:parallelism}
\end{equation}
Despite the possibility of obtaining all possible outputs of a Boolean function in a single query, quantum parallelism cannot be used directly.

Results in quantum computation are obtained via  \emph{measurement} which returns only a limited information about the system. For instance, if a measurement is performed in a quantum state $\ket{\psi} = \alpha_i\ket{\psi_i}$ the result will be $\ket{\psi_i}$ with probability $|\alpha_i|^2$. After measurement state $\ket{\psi}$ collapses to the output obtained and new measurements will result in the same output.

With the definition given above, also adopted by the mainstream quantum literature as~\cite{nielsen:00}, quantum operators are linear operators. In this paper, we suppose the viability of a nonlinear quantum operator $Q$ proposed in~\cite{PhysRevLett.81.3992} whose action is described in Equation~\eqref{eq:nonlin} if at least one $\ket{c_i}$ is equal to $\ket{1}$ otherwise its action is described in Equation~\eqref{eq:nonlin2}.
\begin{equation}
Q \left(\sum_i \ket{\psi_i}\ket{c_i} \right)= \left(\sum_i \ket{\psi_i}\right)\ket{1}
\label{eq:nonlin}
\end{equation}
\begin{equation}
Q \left(\sum_i \ket{\psi_i}\ket{c_i}\right) = \left(\sum_i \ket{\psi_i}\right)\ket{0}
\label{eq:nonlin2}
\end{equation}
The speedup obtained by the application of non-linear operators have been associated with unphysical effects, however in~\cite{czachor1998remarks,czachor1998notes} it is presented a version of this non linear quantum operator free of unphysical influences.

\section{Classical and quantum weightless neural networks}

This work deals with quantum weightless neural networks. Weightless Neural Networks (WNN) are neural networks without weights associated in their connections where the information is stored in a look~up table. The first model of WNN named RAM has been proposed in \cite{Aleksander1966}, since then several neural networks models have been proposed, for instance the Probabilistic Logic Neuron (PLN), the Multi-valued Probabilistic Logic Neuron (MPLN), Goal Seeking Neuron (GSN) and the quantum RAM neuron (qRAM). 

A weightless neuron with $n$ input values has a memory with $2^n$ addressable positions. The learning procedure of a WNN does not require differential calculus or any  complex mathematical calculations. The learning procedure is performed by writing in the look~up table. This learning strategy is faster than techniques based in gradient descendant methods and are suitable to implementation in conventional digital hardware.

     Several models of weightless neural networks are described  on~\cite{Ludermir1999}. In this paper we deal with the qRAM neural network. The qRAM neuron is based in the simplest weightless model, the RAM neuron. Besides its simplicity RAM neurons can be trained very rapidly. Some applications of RAM and RAM based neurons in real world problems are described e.g. in~\cite{Staffa2014,Cardoso2014,Carvalho2014,DeSouza2009}. For a recent review see \cite{esann:2014:tutorial}. In~\cite{Staffa2014, Carvalho2014} a WiSARD system is used to track moving objects or human beings, in~\cite{Cardoso2014} a WiSARD clustering version is proposed to perform credit analysis and in~\cite{DeSouza2009} a VG-RAM weightless neural network is used to perform multi-label text categorisation.

    \subsection{RAM Node}
    A RAM neuron with $n$ inputs has a memory $C$ with $2^n$  addressable positions. Each memory position of a RAM neuron stores a Boolean value and its address is a Boolean string in $\{0,1\}^n$ also called Boolean vector. When a RAM neuron receives a Boolean vector $x=x_1\cdots x_n$ as input it will produce the output $C[x]$. Learning in the qRAM node is very simple and can be accomplished updating the bits in memory positions for each one of the patterns in the training set. 
   
   Architectures of weightless neural networks are weekly connected as a consequence of the limited number of neurons inputs. Two common architectures are pyramidal where the output of a neuron in one layer is connected with a single neuron in the next layer or with only one layer where the neural network output is the sum of each neuron output. 
    
    \subsection{Quantum neural networks}
    
    The notion of quantum neural networks has been proposed on several occasions~\cite{panella:09,ventura:04,Behrman}. In~\cite{panella:09, ventura:04} quantum neural models are pure abstract mathematical devices and in~\cite{Behrman} quantum neural networks are described as a physical device. In this paper we follow the first approach where a neural network is a mathematical model. It is also possible to classify quantum neural networks models as either quantum neural model~\cite{Andrecut2002,panellaneurofuzzy, panella:09,ventura:04,Behrman,daSilva:12} or quantum inspired model~\cite{Li2013,kouda:05}. Quantum inspired models are classical models of computation that uses ideas from quantum computing. Implementation of the quantum weightless neural network mathematically described in this paper requires a real quantum computer. Recent reviews on quantum neural networks can be found in~\cite{Schuldquest,Altaiskycurrent}.
    
    Models of quantum weightless neural networks are proposed or analysed in~\cite{oliveira:08,Oliveira2009,daSilva:12, DaSilva2012}. Quantum weightless neural networks' models are first proposed in~\cite{oliveira:08}, in~\cite{daSilva:10a} a quantum version of the RAM neuron based on an associative quantum memory is presented and in~\cite{daSilva:12} the qRAM neuron and a learning algorithm for quantum weightless neural networks are presented. 
    
    Learning algorithms for quantum neural networks are also proposed in~\cite{panella:09, ventura:04} where a superposition of neural networks with a fixed architecture is created and a quantum search is performed to recover the best neural network architecture. In this paper we propose a variation of this methodology to train  quantum weightless neural networks. In our training strategy, weightless neural networks with different architectures are in a superposition. The neural network model used in this learning methodology is the qRAM neural network.

    \subsection{qRAM Node}
    
The qRAM neuron is a quantum version of the RAM neuron. As in classical case, a $n$ input qRAM neuron has a quantum memory with $2^n$ memory positions. The content of the qRAM memory cannot be directly stored because a measurement of the output could destroy the information stored in the qRAM memory. We store quantum bits in the computational basis named selectors and apply one quantum operator $A$ to obtain the stored qubit.

The $A$ operator used in the qRAM is the control $X$ operator described in Equation~\ref{eq:CNOT}. With this operator a quantum RAM neuron can be described as in Definition~\ref{def:qram}, where memory contents are stored in quantum register selectors.

    \begin{equation}
    \begin{array}{lr}
    A = \begin{pmatrix}
    I & 0 \\
    0 & X 
    \end{pmatrix}
    &
    \begin{array}{l}
    \mbox{where}\\
    A\ket{00} = \ket{0}I\ket{0}\\
    A\ket{10} = \ket{1}X\ket{0}\\
    \end{array}
    \end{array}
    \label{eq:CNOT}
    \end{equation}
    
    \begin{definition}
 A qRAM node with $n$ inputs is represented by the operator $\textsf{N}$ 
described in Equation~\eqref{eq:N}. The inputs, selectors and outputs of 
$\textsf{N}$ are organised in three quantum registers $\ket{i}$ with $n$ qubits, 
$\ket{s}$ with $2^n$  qubits and $\ket{o}$ with 1 qubit. The quantum state 
$\ket{i}$ describe qRAM input, and quantum state $\ket{s}\ket{o}$ describes qRAM 
state.
\label{def:qram}
\end{definition}
\begin{equation}
\textsf{N} = \sum_{i=0}^{2^n-1} \ket{i}_n \bra{i}_n A_{s_i,o}
\label{eq:N}
\end{equation}
    
The qRAM neural network functions exactly as a RAM neural network when the selectors are in the computational basis. For instance, when the quantum register selectors of a qRAM neuron is in the state $\ket{c_{00}c_{01}c_{10}c_{11}}$ and the input $\ket{xy}$ is presented its output is $\ket{c_{xy}}$. The difference between the qRAM and RAM neurons can be observed when the selectors are initialised with a state in superposition. Suppose an initialisation of the quantum register selector with a state in the superposition $\frac{1}{\sqrt{2}}\left( \ket{c_{00}c_{01}c_{10}c_{11}} +  \ket{c'_{00}c'_{01}c'_{10}c'_{11}}\right)$. When the neuron receives an input $\ket{xy}$, the output for each configuration in the superposition will be calculated and the quantum register output will be in the state $\frac{1}{\sqrt{2}}\left(\ket{ c_{xy} } + \ket{ c'_{xy} }\right)$, a sort of parallel execution of the network.

Classical and quantum weightless neurons require a memory (in the classical case) and a number of selectors (in the quantum case) exponential in relation to the number of inputs. To avoid exponential memory requirements,  classical and quantum weightless neural networks use a feed-forward, low connected, pyramidal architecture. A pyramidal, feed-forward neural network with  three two inputs qRAM Nodes is shown in Figure~\ref{fig:qRAMNet1}. A pyramidal qRAM network with $n$ inputs and composed of neurons with two inputs will have $2^{log_2(n)}-1 = n-1$ neurons. Each two input neuron has a memory with 4 selectors than network memory  will need of $4 \cdot (n-1)$ selectors (linear memory size instead of an exponential memory size).

Configuration of a qRAM neural network is realised by the neuron selectors. For instance the configuration of the qRAM neural network in Figure~\ref{fig:qRAMNet1} is the state of quantum registers $\ket{s_1,s_2,s_3}$. For instance, a qRAM network with architecture displayed in Figure~\ref{fig:qRAMNet1} with configuration $\ket{s_1}=\ket{0110}$, $\ket{s_2}=\ket{0110}$ and $\ket{s_3}=\ket{0110}$ can solve the 4 bit parity problem. Superposition of qRAM neural networks with a given architecture can be obtained with the initialisation of qRAM neural configuration with a state in superposition. In Section 5, we explore superposition of qRAM networks in the learning procedure to allow neural network architecture selection.

    \begin{center}
        \begin{figure}
        \center
            \setlength{\unitlength}{0.7mm}
            \begin{picture}(85,50)(0,0)
            \put(20,24){\framebox(15,15)}
            \put(26,31){$\textsf{N}_1$}
            \put(0,35){$i_1$}
            \put(0,30){$i_2$}
            \put(5,25){$s_1$}
            \put(5,35.5){\vector(1,0){10}}
            \put(5,30.5){\vector(1,0){10}}
            \put(10,25.5){\vector(1,0){5}}
            \put(20,5){\framebox(15,15)}
            \put(26,11){$\textsf{N}_2$}
            \put(0,15){$i_3$}
            \put(0,10){$i_4$}
            \put(5,5){$s_2$}
            \put(5,15.5){\vector(1,0){10}}
            \put(5,10.5){\vector(1,0){10}}
            \put(10,5.5){\vector(1,0){5}}
            \put(35,32.5){\line(1,0){10}}
            \put(45,32.5){\line(0,-1){5}}
            \put(45,27.5){\vector(1,0){18}}
            \put(35,12.5){\line(1,0){10}}
            \put(45,12.5){\line(0,1){10}}
            \put(45,22.5){\vector(1,0){18}}
            \put(65,15){\framebox(15,15)}
            \put(71,21){$\textsf{N}_3$}
            \put(53,15){$s_3$}
            \put(58,15.5){\vector(1,0){5}}
            \put(80,22.5){\vector(1,0){10}}
            \end{picture}
            \caption{ qRAM Neural Network of 2 layers}
            \label{fig:qRAMNet1}
        \end{figure}
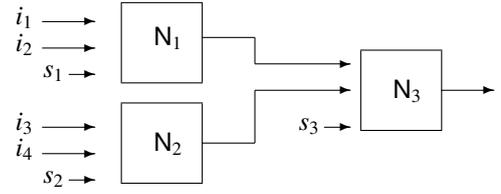
    \end{center}


\section{Non linear quantum learning}
\begin{figure}[ht]%
\begin{center}
\includegraphics[width=0.7\columnwidth]{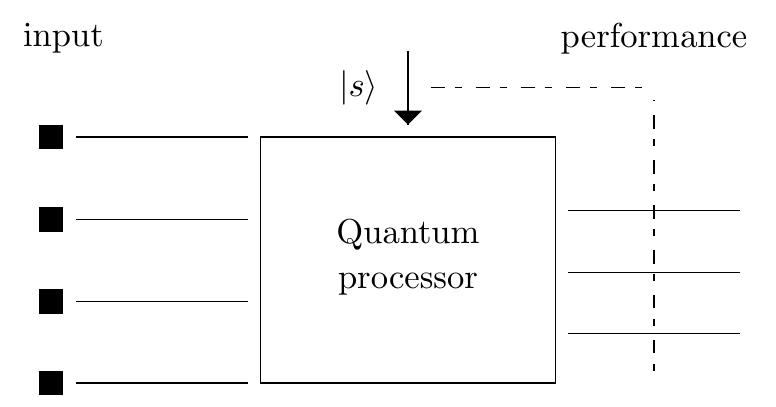}
\end{center}
\caption{Measurement and feedback methodology}%
\label{fig:mf}%
\end{figure}

Nonlinear quantum operators have been used previously~\cite{panella:09, zhou:12}. In this section we show how to train a weightless neural network with a nonlinear quantum algorithm. The proposed algorithm is based on a strategy proposed in~\cite{Gammelmark:09}, where the learning procedure is performed by measurement and feedback. Figure~\ref{fig:mf} illustrates how the measurement and feedback strategy works. The input is presented to a controlled quantum operator named quantum processor, and the result of a measurement performed in the output registers is used to update qubits in the control quantum register. The procedure is repeated until the control qubits $\ket{s}$ are set to some desired value.

\begin{figure}%
\resizebox{\columnwidth}{!}{
\input{framework}
}
\caption{Action of quantum processor in Figure~\ref{fig:mf} when the selector quantum register of a qRAM weightless neural network with fixed architecture is a superposition of quantum states}%
\label{fig:qp}%
\end{figure}
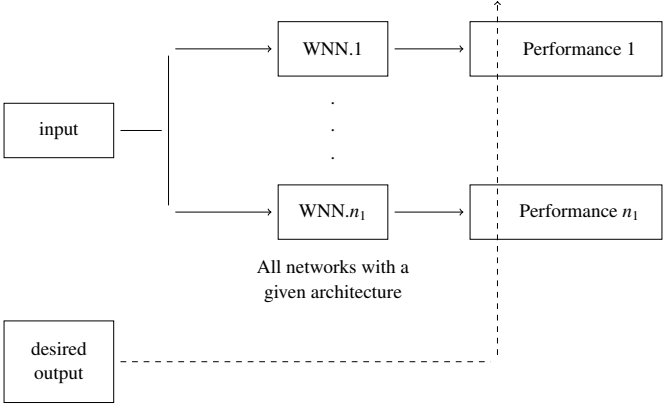

The quantum processor in our learning strategy will be a qRAM weightless neural network with a fixed architecture. 
This quantum weightless neural network can have any number of layers and neurons and must have a feed-forward architecture. 
Patterns selected from a training set will be presented to several neural networks in parallel. 
This step cannot be efficiently performed in a classical computer, but it can be performed in a quantum computer using quantum parallelism. 
Operation performed by the quantum processor is described in Figure~\ref{fig:qp} where each pattern $x$ is presented to  all qRAM network configurations represented in the quantum register $\ket{s}$ and the performance quantum register is updated to indicate if the neural network output is equal to the desired output $d(x)$. After the presentation of all patterns in the training set all pairs of neural network configuration and its respective performance will be in  superposition.

Control qubits of the quantum processor in the measurement and feedback strategy are selectors of the qRAM neural network. 
In the $k$th iteration of the measurement and feedback methodology a non-linear quantum operator and a measurement are performed to determine the $k$th quantum bit of the selectors quantum register $\ket{s}$. 
After all iterations, the quantum register $\ket{s}$ will hold a qRAM configuration with performance greater than or equal to a given threshold $\theta$  for given training set (if exists).

\begin{algorithm}[ht]
\caption{Learning algorithm}
\label{alg:la}
\For{$k=1$ to $n_s$ \label{line:for1}}{
Set input quantum register $\ket{i}$ to $\ket{0}$ \label{line:initinput}\\
Set the $n_s-k+1$ last qubits in quantum register $\ket{s}$ to $H\ket{0}$ \label{line:setselect}\\
Set output quantum register to $\ket{0}$ \label{line:initoutput}\\
Set performance quantum register to $\ket{0}$ \label{line:initper}\\
Set objective quantum register to $\ket{0}$ \label{line:initobj}\\
				\For{each pattern $x \in$ training set \label{line:for2}}{
			Set quantum register $\ket{i}$ to $\ket{x}$ and quantum 
register $\ket{d}$ to $\ket{d(x)}$\label{line:loadpattern}\\
			Allow the qRAM network to produce it output in quantum register $\ket{o}$ \label{line:run}\\
			\If{$\ket{o} = \ket{d(x)}$ \label{line:if}}{add 1 into quantum register $\ket{\perf}$ \label{line:endif}}
		Remove $\ket{x}$ and $\ket{d(x)}$ of quantum registers $\ket{i}$ and $\ket{d}$ \label{line:remove}\\
		}

	\For{$l = 0$ to 1 \label{line:for3}}{
		Set quantum register objective to $\ket{1}$ if $k$th quantum bit in neuron representation 
		is equal to $l$ and performance is greater than a given threshold $\theta$. \label{line:obj}\\
		Apply the non-linear quantum operator NQ to quantum register objective. \label{line:nonlin}\\
		\If{$\ket{objective} = \ket{1}$ \label{line:if2}}{
		Perform a measurement in all quantum register\\
		Set $k$th bit of quantum register selectors to $l$	\label{line:endif2}
		}
	}
	\label{line:endfor3}
}
\end{algorithm}

Algorithm~\ref{alg:la} presents the proposed learning strategy. It requires six quantum registers. 
Input quantum register $\ket{i}$ used to present patterns from the training set to the qRAM network. The free parameters or selectors quantum register $\ket{s}$ used to store qRAM neural network configuration. The output quantum register $\ket{o}$ used to store the qRAM neural network output, the desired output quantum register $\ket{d}$, performance quantum register $\ket{\perf}$ used to store the performance of each classifier in the superposition. And the objective quantum register $\ket{obj}$ used to mark  configurations with desired performance. A configuration of the weightless neuron during the execution of Algorithm~\ref{alg:la} will be represented using the quantum state $\ket{\psi}$ described in Equation~\eqref{eq:qr}.
\begin{equation}
\ket{\psi} = \ket{i}\ket{s}\ket{o}\ket{d}\ket{\perf}\ket{obj}
\label{eq:qr}
\end{equation}

The for loop starting in line~\ref{line:for1} will be repeated $n_s$ times, where $n_s$ is the number of quantum bits in quantum register $\ket{s}$. At the end of the $k$th iteration a non-linear quantum operator is performed to determine the $k$th bit $l_k$ of the quantum register $\ket{s}$.

Steps~\ref{line:initinput},~\ref{line:initoutput},~\ref{line:initper},~\ref{line:initobj} initialise quantum registers input, output, performance and objective. Step~\ref{line:setselect} of Algorithm~\ref{alg:la} initialises selector quantum register. After this step, the state of quantum registers $\ket{s}$ is described in Equation~\eqref{eq:init}, where the value of first $k$ qubits $l_i$ were determined in $i$th iteration of the for loop and the last $n_s-k$ qubits are initialised with $H\ket{0}$ state. 
\begin{equation}
\ket{s} = \left(\frac{1}{\sqrt{2}}\right)^{n_s-k+1}\ket{l_1 \cdots l_{k-1}} \left(\ket{0}+\ket{1}\right)^{\otimes(n_s-k+1)}
\label{eq:init}
\end{equation}

The for loop starting in line~\ref{line:for2} performs the quantum processor operation. It calculates the performance of all configurations in the superposition for the given architecture simultaneously due to principle of quantum parallelism. Step~\ref{line:loadpattern} initialises quantum register input with a pattern $x$ from the data-set, and desired output quantum register with the desired output of $x$ named $d(x)$. These initialisation steps can be performed by unitary operators controlled by a classical system using the classical representation of $x$ and $d(x)$. Step~\ref{line:run} runs the qRAM neural network and its output quantum register is set to the calculated output $y(x,s)$ for pattern $x$ with neural network configuration $s$. Steps~\ref{line:if} to~\ref{line:endif} adds 1 to quantum register performance if $y(x,s)$ is equal to $d(x)$. After these steps, description of state $\ket{\psi}$ is presented in Equation~\eqref{eq:step4}, where state $\ket{s}$ is described in Equation~\eqref{eq:init} and $\ket{\perf(x,s)}$ is the performance of the neural network with selectors $s$ after reading the input $x$.
\begin{equation}
\ket{\psi} = \ket{x}\ket{s}\ket{y(x,s)}\ket{d(x)}\ket{\perf(x,s)}\ket{0}
\label{eq:step4}
\end{equation}

Step~\ref{line:remove} removes $\ket{x}$ and $\ket{d(x)}$ of quantum registers $\ket{i}$ and $\ket{d}$ performing the inverse operation of Step~\ref{line:loadpattern}. After the execution of the for loop starting in line~\ref{line:for2} the performance of each classifier $\ket{\perf(s)}$ will be in superposition with its representation $\ket{s}$.

The for loop starting in line~\ref{line:for3} performs the measurement and feedback. 
An exhaustive non-linear quantum search is performed to determine the value of the $k$th bit in quantum state $\ket{s}$.
Step 16 sets the quantum register $\ket{obj}$ to $\ket{1}$ if $\perf(s) =\theta$ and $k=l$. This step can be performed by a unitary controlled operator $U_g$ that flips objective quantum register if and only if $\perf(x,s) \geq \theta$ and $k=l$. 
After Step 16 the state of quantum registers $\ket{s}$, $\ket{\perf}$ and $\ket{obj}$ is described in Equation~\eqref{eq:f}, 
where $\delta_{ms,l,\perf(ms)}$   is equal to 1 if $\perf(s) \geq \theta$ and the $k$th quantum bit in $\ket{s}$ is equal to $l$.
\begin{equation}
\ket{s,\perf,obj}=\ket{s,\perf(s),\delta_{s,l,\perf(s)}}
\label{eq:f}
\end{equation}

All previous steps can be performed utilising only linear quantum operators. 
Step~\ref{line:nonlin} applies the non-linear quantum operator proposed in~\cite{PhysRevLett.81.3992} to the objective quantum register. 
The objective quantum register will be changed to the basis state $\ket{1}$ if there is at least one configuration in the superposition with objective equal to one. 
In this case, Steps~\ref{line:if2} to~\ref{line:endif2} performs a measurement in state $\ket{\psi}$ and changes the $k$th quantum bit in quantum register $\ket{s}$ to $l$.

The computational cost of Algorithm 1 depends on the number of patterns in the training set $n_t$ and on the number of qubits used in selector quantum register $n_s$. The for loop starting in line 1 will be repeated $n_s$ times. Steps~\ref{line:initinput} to~\ref{line:initobj} have constant computational time. For loop in lines~\ref{line:for2} to 13 will be repeated $n_t$ times and each inner line has constant computational cost. For loop in lines~\ref{line:for3} to \ref{line:endfor3} does not depend on $n_t$ and $n_s$ and it has a constant computational cost. In this way the overall cost of the Algorithm 1 is $O(n_t \cdot n_s )$. Then Algorithm 1 has polynomial time in relation to the number of qubits used to represent the qRAM neural network selectors and the number of patterns in the training set.

A concrete example of Algorithm 1 execution is presented to illustrate its functionality. 
Without loss of generality we use a qRAM neural network composed by only one neuron with two inputs to learn the 2-bit XOR toy problem described in Equation~\eqref{eq:xor}. For this problem, quantum register input needs two qubits, quantum register selectors has 4 qubits, quantum register output needs 1 qubit, quantum register performance has 3 qubits and quantum register objective has 1 qubit.

\begin{equation}
T=\left\{\left(\ket{00},\ket{0}\right),\left(\ket{01},\ket{1}\right),\left(\ket{10},\ket{1}\right),\left(\ket{11},\ket{0}\right)\right\}
\label{eq:xor}
\end{equation}

In Steps 2, 4, 5 and 6 bits in quantum registers input, output, performance and objective are initialised with the quantum state $\ket{0}$. The number of quantum bits in $\ket{s}$ quantum register is equal to 4 and in the first iteration $n_s-k+1$ is also equal to 4, then all four qubits in quantum register $\ket{s}$ are initialised with the state $H\ket{0}$. After these initialisation steps, neural network configuration $\ket{\psi}$ is described in Equation~\eqref{eq:6}.
\begin{equation}
\begin{split}
\ket{\psi} = \frac{1}{4}\ket{00}\left(\ket{0}+\ket{1}\right)^{\otimes 4}\ket{0}\ket{0}\ket{000}\ket{0} = \\
\frac{1}{4}\sum_{j\in \left\{0,1\right\}^4}\ket{00}\ket{j}\ket{0}\ket{0}\ket{000}\ket{0}
\end{split}
\label{eq:6}
\end{equation}

Suppose that in the first iteration of the for loop starting in line~\ref{line:for1} $x$ assumes value $\ket{01}$ and $d(x)$ is $\ket{1}$. Step~\ref{line:loadpattern} initialises pattern and desired output quantum register respectively to $\ket{01}$ and $\ket{1}$. This initialisation can be performed through CNOT operators applied to $\ket{\psi}$ resulting in state $\ket{\psi_1}$ described in Equation~\eqref{eq:7}.

\begin{equation}
\frac{1}{4}\sum_{j\in \left\{0,1\right\}^4}\ket{01}\ket{j}\ket{0}\ket{1}\ket{000}\ket{0}
\label{eq:7}
\end{equation}

Step~\ref{line:run} runs the neural network and this output is calculated in quantum register $\ket{o}$. After this step we obtain the state $\ket{\psi_2}$ described in Equation~\eqref{eq:8}, where $j_1$ is the qubit in memory position 01 and $\delta_{j_1,1}=1$ if and only if $j_1=1$.
\begin{equation}
\frac{1}{4}\sum_{j\in \left\{0,1\right\}^4}\ket{01}\ket{j}\ket{\delta_{j_1,1}}\ket{1}\ket{000}\ket{0}
\label{eq:8}
\end{equation}

Step~\ref{line:if} to~\ref{line:endif} check if desired output is equal to the calculated output, adding one to the performance quantum register if they are equal.  The resulting state after Step~\ref{line:endif} $\ket{\psi_3}$ is described in Equation~\eqref{eq:9}. These steps can be performed using a unitary operator describing the qRAM neural network and a quantum operator that adds one to the quantum register performance with controls $\ket{o}$ and $\ket{d}$.
\begin{equation}
\begin{split}
\ket{\psi_3} = \frac{1}{4}\left(\ket{01}\ket{0000}\ket{0}\ket{1}\ket{000}\ket{0} \right. \\
+  \ket{01}\ket{0001}\ket{0}\ket{1}\ket{000}\ket{0} 
+  \ket{01}\ket{0001}\ket{0}\ket{1}\ket{000}\ket{0}  \\
+  \ket{01}\ket{0001}\ket{0}\ket{1}\ket{000}\ket{0} 
+  \ket{01}\ket{0001}\ket{0}\ket{1}\ket{000}\ket{0} \\
+  \ket{01}\ket{0001}\ket{0}\ket{1}\ket{000}\ket{0} 
+  \ket{01}\ket{0001}\ket{0}\ket{1}\ket{000}\ket{0} \\
+  \ket{01}\ket{0001}\ket{0}\ket{1}\ket{000}\ket{0} 
+  \ket{01}\ket{0001}\ket{0}\ket{1}\ket{000}\ket{0} \\
+  \ket{01}\ket{0001}\ket{0}\ket{1}\ket{000}\ket{0} 
+  \ket{01}\ket{0001}\ket{0}\ket{1}\ket{000}\ket{0} \\
+  \ket{01}\ket{0001}\ket{0}\ket{1}\ket{000}\ket{0} 
+  \ket{01}\ket{0001}\ket{0}\ket{1}\ket{000}\ket{0} \\
+  \ket{01}\ket{0001}\ket{0}\ket{1}\ket{000}\ket{0} 
+  \ket{01}\ket{0001}\ket{0}\ket{1}\ket{000}\ket{0} \\
+  \ket{01}\ket{0001}\ket{0}\ket{1}\ket{000}\ket{0} 
\end{split}
\label{eq:9}
\end{equation}

Step~\ref{line:remove} removes the values of $\ket{x}$ and $\ket{d(x)}$ from quantum registers $\ket{i}$ and $\ket{d}$ allowing the initialisation of the next for loop iteration. After the for loop last execution only one configuration in superposition, with $\ket{s}=\ket{0110}$, has performance 100\% and the selectors and performance quantum registers are described by quantum state in Equation \eqref{eq:10}, where $\perf(j)<4$ for all $j \neq 0110$.
\begin{equation}
\ket{s,\perf} = \frac{1}{4}\left(\ket{0110}\ket{4}_3 + \sum_{j\in \{0,1\}^4, j\neq 0110}\ket{j}\ket{\perf(j)}\right)
\label{eq:10}
\end{equation}

Setting $\theta$ to 100\%, in the first iteration of the for loop ($l=0$) in line \ref{line:for3}, Step~\ref{line:obj} changes objective register to $\ket{1}$ when the $k$th qubit of $\ket{s}$ is $\ket{0}$ and performance is $\theta$. After Step 15 selectors, performance and objective quantum registers are described in Equation (11).
\begin{equation}
\ket{s,\perf,obj} = \frac{1}{4}\left(\ket{0110}\ket{4}_3\ket{1} + \sum_{j\in \{0,1\}^4, j\neq 0110}\ket{j}\ket{\perf(j)}\ket{0}\right)
\label{eq:11}
\end{equation}

Step~\ref{line:nonlin} applies the nonlinear quantum operator in objective quantum register and the state of selectors, performance and objective quantum registers are described in Equation~\eqref{eq:12}. The nonlinear quantum operator sets the quantum register objective to $\ket{1}$.
\begin{equation}
\ket{s,\perf,obj} = \frac{1}{4}\left(\ket{0110}\ket{4}_3 + \sum_{j\in \{0,1\}^4, j\neq 0110}\ket{j}\ket{\perf(j)}\right)\ket{1}
\label{eq:12}
\end{equation}

Since the objective quantum register is in a base state we can check whether $\ket{obj} = \ket{1}$ with no information loss. In Steps~\ref{line:if2} to~\ref{line:endif2} a measurement is performed in quantum register $\ket{s}$ and the first qubit of $\ket{s}$ is set to $\ket{l_1} = \ket{0}$. This qubit will not be changed in the next iterations.

At the end of the main for loop the selector quantum register $\ket{s}$ will be in the state $\ket{0110}$ and the desired configuration was found. Next section shows how to perform a search in the architecture space of a quantum weightless neural network.

\section{Architecture learning}

The operator $A$ in a  qRAM neural network is known as controlled not operator. In other 
models of quantum weightless neural networks this operator can assume different 
forms. For instance, in \cite{oliveira:08} the $A$ operators of qPLN  are 
represented in computational basis by the quantum operator $A_{qPLN}$ described 
in Equation~\eqref{eq:aqpln}, where $\textsf{U}$ is an arbitrary quantum 
operator
\begin{equation}
\begin{split}
 A_{qPLN} =\ket{00}\bra{00}\otimes \textsf{I} + \ket{01}\bra{01} 
\otimes \textsf{X} + \\ \ket{10}\bra{10}\otimes\textsf{H} + 
\ket{11}\bra{11}\otimes\textsf{U} 
\end{split}
\label{eq:aqpln}
\end{equation}
and  
the $A$ operators of a qMPLN with $n$ qubits in each memory position  are 
represented by the matrix described in Equation~\eqref{eq:aqmpln}, where 
$\qop{U}_{p_k}$ is a rotation operator with angle $p_k$.
\begin{equation}
 A_{qMPLN} = \sum_{k=0}^{n-1}\ket{k}\bra{k}\otimes \qop{U}_{p_k}
 \label{eq:aqmpln}
\end{equation}

These $A$ operators are used to generate the values stored in a specific 
memory position. For instance in the qPLN, instead of storing the qubit 
$\frac{1}{\sqrt{2}}\left(\ket{0} + \ket{1}\right)$, we store the qubits in the computational basis $\ket{10}$ and uses 
the operator $\qop{A}_{qPLN}$ to generate the content $\frac{1}{\sqrt{2}}\left(\ket{0} + \ket{1}\right)$.

\begin{figure}%
\includegraphics[width=\columnwidth]{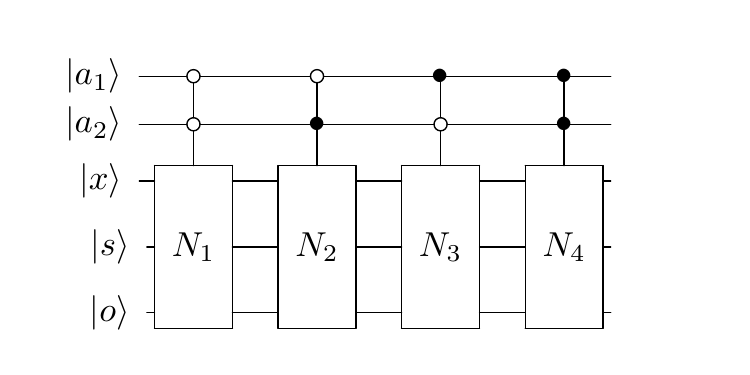}%
\caption{Quantum neuron representing a weightless neural networks with four different architectures}%
\label{fig:qn}%
\end{figure}

 The main idea in this Section is to allow a weightless neural network to store 
the output of a weightless neural network with a given input $x$ and selectors $s$. In this case, the quantum 
version of this weightless neural network will need a matrix $A$ representing 
the weightless neural network to generate the output of the weightless neural network. Then the 
$A$ operators are replaced by operators representing weightless neural networks and selectors 
are replaced by the neural network inputs and selectors. Figure~\ref{fig:qn} illustrates this weightless neuron with two inputs, where $\ket{a_1 a_2}$ are architecture selectors, input pattern $x$ and selectors are combined in one single quantum register and acts as the free parameters of the neuron, and quantum register output is shared by all weightless networks $N_0,N_1,N_2,N_3$.

\begin{figure}%
\resizebox{\columnwidth}{!}{
\input{framework2}
}
\caption{Action of quantum processor in Figure~\ref{fig:mf} when the selector and architecture selector quantum registers of a weightless neuron with some distinct architectures are in a superposition of quantum states}%
\label{fig:fram2}%
\end{figure}
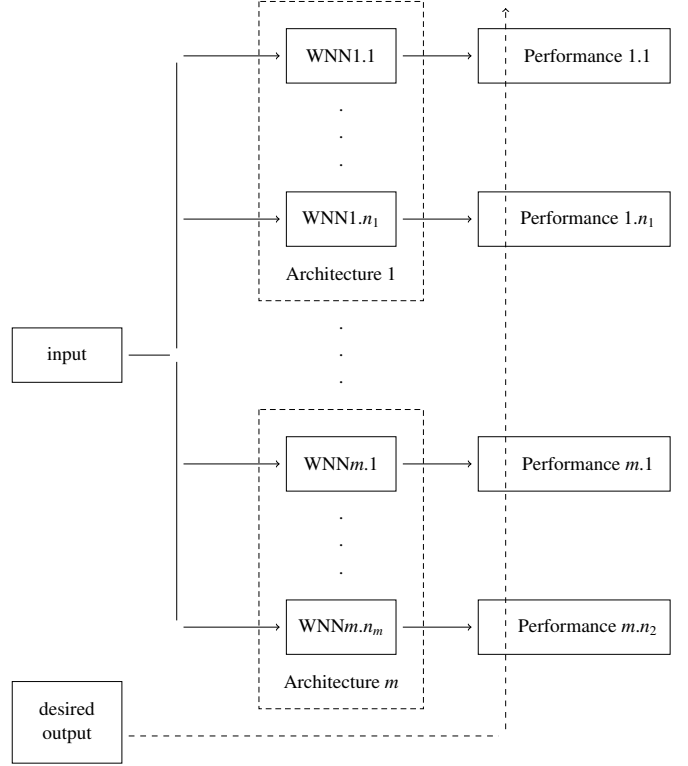

With this quantum neuron the action of the quantum processor in Figure~\ref{fig:mf} can be described by Figure~\ref{fig:fram2}. Initialisation of the architecture selector quantum register with a quantum state in superposition will put different architectures, represented by the doted boxes, into superposition. And the initialisation of selectors quantum registers puts different configurations of each architecture into superposition. Problem of architecture selection  is reduced to the problem of training the weightless neuron in Figure~\ref{fig:qn} where the input is represented by quantum register $\ket{x}$ and selectors are represented by quantum registers $\ket{a,s}$. In this way, Algorithm 1 can be used to learning parameters and architecture simultaneously.

Architecture selection computational time is directly related to computational time of the Algorithm~\ref{alg:la}. Due to the linearity of quantum operators, neurons can share selectors and under supposition that all architectures are pyramidal and low connected then network memory size (or the necessary number of selectors) will be polynomial in relation to the number of neural network inputs.
The cost of architecture selection will be $O\left(n_a+n_s+n_t\right)$, where $n_a$ is the number of architectures, $n_s$ is the number of selectors in the most complex (with more selectors) architecture and $n_t$ is the number of training patterns.

\subsection{Architecture selection with SAL algorithm}
Quantum computers are not yet a reality and we cannot evaluate SAL algorithm in real problems. In this Section we present a concrete example (with low dimensionality) of the SAL algorithm in architecture selection. We
use the artificial dataset described in Table~\ref{tab:artificialDataSet}
obtained in the following way. Two weightless neural network architectures were defined and an exhaustive search was performed to find a dataset in each one architecture can learn the dataset and the other architecture cannot learn the dataset using selectors in the computational basis.

\begin{table}%
\begin{center}
\begin{tabular}{|cccc|c|}\hline
\multicolumn{4}{|c|}{Patterns}   & Class \\ \hline
0 & 0 & 0 & 0 & 1 \\
0 & 0 & 0 & 1 & 1 \\
0 & 0 & 1 & 0 & 0 \\
0 & 0 & 1 & 1 & 1 \\
0 & 1 & 0 & 0 & 1 \\
0 & 1 & 0 & 1 & 1 \\
0 & 1 & 1 & 0 & 0 \\
0 & 1 & 1 & 1 & 1 \\
1 & 0 & 0 & 0 & 1 \\
1 & 0 & 0 & 1 & 1 \\
1 & 0 & 1 & 0 & 0 \\
1 & 0 & 1 & 1 & 1 \\
1 & 1 & 0 & 0 & 0 \\
1 & 1 & 0 & 1 & 1 \\ 
1 & 1 & 1 & 0 & 0 \\
1 & 1 & 1 & 1 & 1 \\ \hline
\end{tabular}
\end{center}
\caption{Simple artificial data set}
\label{tab:artificialDataSet}
\end{table}

The architectures used in the experiment are two layers, pyramidal qRAM weightless neural networks. The first architecture $\qop{N}_0$ has two qRAM neurons each with two inputs in the first layer and one qRAM neuron with two inputs in the second layer. Figure~\ref{fig:qRAMNet1} displays architecture $\qop{N}_0$. The second architecture $\qop{N}_1$ has two qRAM neurons in the first layer where the first neuron has three inputs and the second neuron has one input and the second layer has one qRAM neuron with two inputs.

The first architecture needs of 12 quantum bits for representing selector quantum register, 4 quantum bits for representing input of the first layer, 2 quantum bits to represent the second layer input, and 1 quantum bit to representing the neural network output. In this way, the first architecture representation needs of 19 quantum bits. The second architecture needs of 14 quantum bits for representing selector quantum register and the same number of quantum bits used by the first architecture to represent neurons inputs and network output than the second architecture representation requires 21 quantum bits. 

These two qRAM neural networks are represented in a single circuit with  six quantum registers. Neurons inputs quantum register $\ket{i}$ with 6 quantum bits, selectors quantum register $\ket{s}$ with 14 quantum bits, output quantum register $\ket{o}$ with one qubit and architecture selector quantum register $\ket{a}$ with 1 qubit. Performance quantum register $\ket{\perf}$ with 5 quantum bits. Output quantum register with 1 quantum bit. 

The qRAM neural network with architecture $\qop{N}_0$ uses all qubits in quantum registers selectors, input and output. The qRAM neural network with architecture $\qop{N}_1$ uses all qubits in inputs and output quantum register and uses only 12 qubits in selectors quantum register. The architecture quantum register $\ket{a}$ is used to select the architecture. If $\ket{a}$ is equal to 0 the architecture 1 is used. If $\ket{a}$ is equal to 1 the architecture 2 is used.


After the initialization steps of Algorithm~\ref{alg:la}, the state of quantum registers $\ket{a}\ket{s}\ket{\perf}$ is described in Equation~\eqref{eq:aftinit}, where $\ket{a}$ and $\ket{s}$ are in a superposition with all possible values and the quantum bits in performance quantum register are initialized with $\ket{0}$.
\begin{equation}
\ket{a}\ket{s}\ket{\perf} = (\ket{0}+\ket{1})\sum_{k\in\{0,1\}^{14}}\ket{k}\ket{00000}
\label{eq:aftinit}
\end{equation}
After the datased presetation to the neural network performed in Steps 7 to 14 of Algorithm~\ref{alg:la}, the state of quantum registers $\ket{a}\ket{s}\ket{\perf}$ is described in Equation~\eqref{eq:aftpresentation}, where $\perf(k,N_i)$ is the performance of qRAM neural network with architecture $N_i$ and selectors $\ket{k}$.

\begin{equation}
\begin{split}
\ket{a}\ket{s}\ket{\perf} = \\ \ket{0}\left(\sum_{k\in\{0,1\}^{12}}\ket{k}\qop{H}^{\otimes 2}\ket{00}\right)\ket{\perf(k,N_0)} \\ +\ket{1}\sum_{k\in\{0,1\}^{14}}\ket{k}\ket{\perf(k,N_1)}
\end{split}
\label{eq:aftpresentation}
\end{equation}

$\qop{N}_0$ architecture cannot learn the dataset with 100\% of accuracy and $\qop{N}_1$ can learn the dataset with 100\% of accuracy when its selectors are in the set
\begin{equation}\begin{split}
T = \{\ket{0 1 0 1 0 1 1 1, 0 1, 1 1 0 1}, 
  \ket{0 1 0 1 0 1 1 1, 1 0, 1 1 1 0}, \\
\ket{1 0 1 0 1 0 0 0, 0 1, 0 1 1 1}, 
\ket{1 0 1 0 1 0 0 0, 1 0, 1 0 1 1} \}.\\
 \end{split}\end{equation}
 In the second iteration of for loop starting in line 15, the quantum register objective is set to $\ket{1}$ if and only if the performance is greather than a given threshold $\theta$. Here we use $\theta$ equal to 16 (100\% of accuracy), after this operation the state of quantum registers $\ket{a}\ket{s}\ket{\perf}\ket{obj}$ is described in Equation~\eqref{eq:afttheta}.

\begin{equation}
\begin{split}
\ket{a}\ket{s}\ket{\perf}\ket{obj} =\\ \ket{0}\left(\sum_{k\in\{0,1\}^{12}}\ket{k}\qop{H}^{\otimes 2}\ket{00}\right)\ket{\perf(k,N_0)}\ket{0} \\ +\ket{1}\sum_{k\in\{0,1\}^{14}, k\notin T}\ket{k}\ket{\perf(k,N_1)}\ket{0} \\ +\ket{1}\sum_{k\in T}\ket{k}\ket{\perf(k,N_1)}\ket{1}
\end{split}
\label{eq:afttheta}
\end{equation}

Step 17 applies the nonlinear quantum operator and the resultant state of quantum registers $\ket{a}\ket{s}\ket{\perf}\ket{obj}$ is described in Equation~\eqref{eq:aftnonlinear}, where a measurement can be performed and the architecture register will be in state $\ket{1}$ and the architecture $\qop{N}_1$ was chosen.

\begin{equation}
\begin{split}
\ket{a}\ket{s}\ket{\perf}\ket{obj} =  \ket{1}\sum_{k\in T}\ket{k}\ket{\perf(k,N_1)}\ket{1}
\end{split}
\label{eq:aftnonlinear}
\end{equation}

\subsection{Discussion}

We proposed a methodology to select quantum neural network parameters and architecture using a quantum weightless neural networks in polynomial time in relation to the number of training patterns, architectures and neural network free parameters. The proposed algorithm, named Superposition based Architecture Learning (SAL), performs a non-linear global search in the space of weightless neural networks parameters and for a given data set returns a classifier with a desired performance $\theta$ or returns that there is no classifier otherwise. 

A classical polynomial time algorithm to perform neural network architecture selection is not known. Classical techniques used to perform architecture selection are heuristics that do not guarantee to find an exact solution. Some strategies used to find near optimal neural networks architectures or parameters are evolutionary algorithms~\cite{Almeida2010} and meta-learning~\cite{Miranda201427}. Running time of evolutionary algorithms used in architecture selection are displayed in~\cite{Miranda201427} and even in benchmark problems the running time of these classical strategies can vary from 3 to 400 minutes.

 In the application of the SAL algorithm to perform architecture selection, if there is a solution in the space search then the solution will be found in polynomial time. SAL algorithm puts all neural network configurations with some architectures in superposition, the performance is calculated and a nonlinear operator is used to recover the configuration and architecture with desired performance. SAL algorithm is the first algorithm to perform quantum weightless neural network architecture selection in polynomial time in relation to the number of patterns, architectures.

Superposition principle allows the evaluation of neural networks architectures in a way that is not possible in classical neural networks. In a classical neural network the architecture evaluation is biased by a choice of neural network parameters. In SAL algorithm all neural network parameters are initialized with all parameters in superposition allowing the evaluation of neural network architecture without the bias of a given set of parameters.

The gain in computational time of the proposed strategy is a result of the use of non-linear quantum operator proposed in~\cite{PhysRevLett.81.3992}. Despite non-linear quantum computing has been used in several works, there still remains some controversy with some authors claiming that non linear quantum operators are not physically realisable~\cite{PhysRevLett.81.3992} while other researchers claiming otherwise~\cite{czachor1998remarks}.

Even if non-linear quantum operators do not become a reality, the proposed learning algorithm furnishes a framework for the development of linear quantum algorithms to perform neural network architecture selection. The proposed idea is to define a quantum weightless neural network such that its memory positions store configurations of neural networks with different architectures.


\section{Conclusion}

For some problems there are quantum algorithms which are asymptotically faster than the known classical algorithms~\cite{grover:96,shor:97,trugenberger:02}.  In this paper, we defined a quantum Superposition based Architecture Learning algorithm for weightless neural networks that finds architecture and parameters with polynomial time in relation to the number of training 	patterns, architectures and the size of the selectors quantum register.  The proposed algorithm used the quantum superposition principle and a nonlinear quantum operator. 

A linear version of the proposed algorithm is challenging research topic which is the subject of on going work. This linear version should be a quantum probabilistic algorithm, because the problem of training a weightless neural networks is a NP-complete problem. One could use the quantum processor to create a superposition of weightless neural networks with different architectures and to perform classical learning steps in these neural networks in superposition before performing the measurement and feedback.

Quantum weightless neural networks proposed in~\cite{oliveira:08} are generalisation of the classical models based on a classical RAM memory. Another possible future work is the analysis of quantum memories~\cite{altaiskymemory,ventura:98} for the development of weightless neural networks models. These quantum memories has an exponential gain in memory capacity when compared with classical memories.

\section*{Acknowledgements}
This work is supported by research grants from CNPq, CAPES and FACEPE (Brazilian research agencies).

  \bibliographystyle{elsarticle-num} 
  \bibliography{bibliografia1}

\end{document}

%% file: framework.tex
\begin{tikzpicture}

\draw  (10,10) rectangle (12,11);
\node at (11,10.5) {WNN.1};
\node at (11,9.5) {.};
\node at (11,9) {.};
\node at (11,8.5) {.};
\draw  (10,8) rectangle (12,7);
\node at (11,7.5) {WNN.$n_1$};
\node at (11,6.5) {All networks with a };
\node at (11,6) {given architecture};

\draw  (5,9.5) rectangle (7,8.5);
\node at (6,9) {input};
\node (v1) at (7,9) {};
\node (v2) at (8,9) {};
\node (v5) at (8,7.5) {};
\node (v3) at (8,10.5) {};
\node (v4) at (10,10.5) {};
\node (v6) at (10,7.5) {};

\draw  (v1) edge (v2);
\draw  (v5) edge (v3);
\draw  (v3) edge[->] (v4);
\draw  (v5) edge [->](v6);

\node (v11) at (12,10.5) {};
\node (v12) at (13.5,10.5) {};
\node (v13) at (12,7.5) {};
\node (v14) at (13.5,7.5) {};

\draw  (v11) edge[->] (v12);
\draw  (v13) edge[->] (v14);
\draw  (5,4.0) rectangle (7,5.5);
\node at (6,5) {desired};
\node at (6,4.5) {output};
\node (v19) at (7,4.75) {};
\node (v20) at (14,4.75) {};
\node (v21) at (14,11.5) {};
\draw [dashed] (v19) edge (v20);
\draw [dashed] (v20) edge[->] (v21);
\draw  (13.5,11) rectangle (17,10);
\node at (15.5,10.5) {Performance 1};
\draw  (13.5,8) rectangle (17,7);
\node at (15.5,7.5) {Performance $n_1$};

\end{tikzpicture}

%% file: framework2.tex
\begin{tikzpicture}

\draw  (10,10) rectangle (12,11);
\node at (11,10.5) {WNN1.1};
\node at (11,9.5) {.};
\node at (11,9) {.};
\node at (11,8.5) {.};
\draw  (10,8) rectangle (12,7);
\node at (11,7.5) {WNN1.$n_1$};
\node at (11,6.5) {Architecture 1};
\draw [densely dashed] (9.5,11.5) rectangle (12.5,6);

\node at (11,5.5) {.};
\node at (11,5.0) {.};
\node at (11,4.5) {.};

\draw  (10,2.5) rectangle (12,3.5);
\node at (11,3) {WNN$m$.1};
\node at (11,2) {.};
\node at (11,1.5) {.};
\node at (11,1) {.};
\draw  (10,0.5) rectangle (12,-0.5);
\node at (11,0) {WNN$m$.$n_m$};
\node at (11,-1) {Architecture $m$};
\draw [densely dashed] (9.5,4) rectangle (12.5,-1.5);
\draw  (5,5.5) rectangle (7,4.5);
\node at (6,5) {input};
\node (v1) at (7,5) {};
\node (v2) at (8,5) {};
\node (v5) at (8,7.5) {};
\node (v3) at (8,10.5) {};
\node (v4) at (10,10.5) {};
\node (v6) at (10,7.5) {};
\node (v9) at (8,3) {};
\node (v7) at (8,0) {};
\node (v8) at (10,0) {};
\node (v10) at (10,3) {};
\draw  (v1) edge (v2);
\draw  (v2) edge (v3);
\draw  (v3) edge[->] (v4);
\draw  (v5) edge [->](v6);
\draw  (v2) edge (v7);
\draw  (v7) edge[->] (v8);
\draw  (v9) edge[->] (v10);
\node (v11) at (12,10.5) {};
\node (v12) at (13.5,10.5) {};
\node (v13) at (12,7.5) {};
\node (v14) at (13.5,7.5) {};
\node (v15) at (12,3) {};
\node (v16) at (13.5,3) {};
\node (v17) at (12,0) {};
\node (v18) at (13.5,0) {};
\draw  (v11) edge[->] (v12);
\draw  (v13) edge[->] (v14);
\draw  (v15) edge[->] (v16);
\draw  (v17) edge[->] (v18);
\draw  (5,-1) rectangle (7,-2.5);
\node at (6,-1.5) {desired};
\node at (6,-2) {output};
\draw  (13.5,11) rectangle (17,10);
\node at (15.5,10.5) {Performance 1.1};
\draw  (13.5,8) rectangle (17,7);
\node at (15.5,7.5) {Performance $1$.$n_1$};
\draw  (13.5,3.5) rectangle (17,2.5);
\node at (15.5,3) {Performance $m$.$1$};
\draw  (13.5,0.5) rectangle (17,-0.5);
\node at (15.5,0) {Performance $m$.$n_2$};
\node (v19) at (7,-2) {};
\node (v20) at (14,-2) {};
\node (v21) at (14,11.5) {};
\draw [dashed] (v19) edge (v20);
\draw [dashed] (v20) edge[->] (v21);
\end{tikzpicture}